\documentclass[prl,aps,showpacs,byrevtex,onecolumn]{revtex4}
\usepackage{graphicx}
\usepackage{amsmath}
\usepackage{amsfonts}

\begin{document}

\title{Quantum Hamiltonian diagonalization and Equations of Motion with
Berry Phase Corrections}
\author{Pierre Gosselin$^{1}$, Alain B\'{e}rard$^{2}$ and Herv\'{e} Mohrbach$
^{2}$}

\address{$^1$Institut Fourier, UMR
5582 CNRS-UJF, UFR de Math\'ematiques, Universit\'e Grenoble I,
BP74, 38402 Saint Martin
d'H\`eres, Cedex, France \\
$^2$Laboratoire de Physique Mol\'eculaire et des Collisions,
ICPMB-FR CNRS 2843, \\ Universit\'e Paul Verlaine-Metz, 57078
Metz Cedex 3, France\\
$^3$D\'epartement de Physique, Universit\'e Mouloud Mammeri -BP 17, Tizi Ouzou, Algerie
}


\begin{abstract}
It has been recently found that the equations of motion of several
semiclassical systems must take into account anomalous velocity terms
arising from Berry phase contributions. Those terms are for instance
responsible for the spin Hall effect in semiconductors or the gravitational
birefringence of photons propagating in a static gravitational field.
Intensive ongoing research on this subject seems to indicate that actually a
broad class of quantum systems might have their dynamics affected by Berry
phase terms. In this article we review the implication of a new
diagonalization method for generic matrix valued Hamiltonians based on a
formal expansion in power of $\hbar$. In this approach both the diagonal
energy operator and dynamical operators which depend on Berry phase terms
and thus form a noncommutative algebra, can be expanded in power series in $
\hbar$. Focusing on the semiclassical approximation, we will see that a
large class of quantum systems, ranging from relativistic Dirac particles in
strong external fields to Bloch electrons in solids have their dynamics
radically modified by Berry terms.
\end{abstract}

\maketitle

\section{Introduction}

Since the seminal work of Berry \cite{BERRY}, the notion of Berry phase has
found several applications in branches of quantum physics such as atomic and
molecular physics, optics and gauge theories. Most studies focus on the
geometric phase that a wave function acquires when a quantum mechanical
system has an adiabatic evolution. Yet, the Berry phase in momentum space
has recently found unexpected applications in several fields. For instance,
in spintronics, such a term is responsible for a transverse dissipationless
spin-current in semiconductors in the presence of electric fields \cite%
{MURAKAMI1}. In optics, it has been recently found that a monopole in
momentum space causes the gravitational birefringence of photons in a static
gravitational fields \cite{PHOTONPIERRE}. Both effects are two
manifestations of the spin Hall effect which can be interpreted at the
semiclassical level as due to the influence of Berry curvatures on the
semiclassical equations of motion of spinning particles \cite{ALAIN}.
Similarly, a new set of semiclassical equations with a Berry phase
correction was proposed to account for the semiclassical dynamics of
electrons in magnetic Bloch bands \cite{NIU1}\cite{PIERREbloch}. In a more
exotic application, intrinsic Berry phase effects in the particle dynamics
of the doubly special relativity theory was recently described \cite%
{SUBIRDIRAC}.

In the above cited examples, the semiclassical equations of motion with
Berry phase corrections (anomalous velocity terms) can be derived in a
representation where the Hamiltonian is diagonalized at the semiclassical
order. It was indeed shown that the semiclassical diagonalization results in
an effective energy operator with Berry phase corrections as well as
noncommutative covariant coordinates and momentum operators. These dynamical
operators being corrected by Berry terms \cite{PIERREGENERAL}, this leads
directly to these new Berry effects. This is another illustration of the
fact that the physical content of quantum systems is most often best
revealed in the representation where the Hamiltonian is diagonal. The
paradigmatic example is provided by the Foldy-Wouthuysen (FW) representation
of the Dirac Hamiltonian for relativistic particles interacting with an
external electromagnetic field. In this representation the positive and
negative energy states are separately represented and the non-relativistic
Pauli-Hamiltonian is obtained \cite{FOLDY}. Actually even if several exact
FW transformations have been found for some definite classes of potentials 
\cite{ERIKSEN}\cite{NIKITIN}\cite{SILENKO1}, the diagonalization of matrix
valued Hamiltonian is, in general, a difficult mathematical problem
requiring some approximations, essentially a perturbation expansion in weak
fields. To overcome this limitation we have recently proposed a new method
based on a formal expansion in powers of the Planck constant $\hbar$ \cite{SERIESPIERRE} 
which is not restricted to Dirac Hamiltonians but also
applicable to a large class of quantum systems. It is worth mentioning that
recently a variant of the FW transformation valid for strong fields and
based also on an expansion in $\hbar$ of the Dirac Hamiltonian was presented 
\cite{SILENKO}. The\ main advantage of the diagonalization procedure of \cite{SERIESPIERRE} 
is that it embraces several different physical systems
ranging from Bloch electrons in solid to Dirac particles interacting with
any type of external fields (for instance in refs. \cite{PHOTONPIERRE}\cite{PIERREelectrongravit} 
electrons and photons in a static gravitational field were considered).

In this paper we review the recursive diagonalization procedure of ref. \cite{SERIESPIERRE} 
from which can deduce the expressions of the semiclassical
energy and of the dynamical operators. We then consider, as a physical
application, an electron in a magnetic Bloch band and the spin hall effect
of light in a gravitational field

\section{Recursive diagonalization of quantum Hamiltonian}

In this section we consider a quantum mechanical system whose state space is
a tensor product $L^{2}\left( \mathcal{R}^{3}\right) \otimes V$\ with $V$\
some internal space. In other words, the Hamiltonian of this system can be
written as a matrix $H_{0}\left( \mathbf{P,R}\right) $\ of size $\dim V$\
whose elements are operators depending on a couple of canonical variables $
\mathbf{P}$\ and $\mathbf{R}$, the archetype example being the Dirac
Hamiltonian with $V=C^{4}$. In \cite{SERIESPIERRE} we found a
diagonalization process for this matrix valued quantum Hamiltonian $
H_{0}\left( \mathbf{P,R}\right) $\ recursively as a series expansion in
powers of $\hbar $ which\ gives the quantum corrections to the diagonalized
Hamiltonian with respect to the classical situation $\hbar =0$. For example,
the first order correction in $\hbar $\ corresponds to the semiclassical
approximation. In this approach we derived the $\hbar $\ expansion
recursively in the following way. The Planck constant $\hbar $\ is formally
promoted to a dynamical parameter $\alpha $\ in order to establish a
differential equation connecting the two diagonalized Hamiltonians at $\hbar
=\alpha $\ and $\hbar =\alpha +d\alpha $. The integration of this
differential equation allows then the recursive determination of the
different terms in the expansion of the diagonalized Hamiltonian in powers
of $\alpha $. To start with, consider the diagonalization at the scale $
\alpha $ 
\begin{equation}
U_{\alpha }\left( \mathbf{P},\mathbf{R}\right) H_{0}\left( \mathbf{P,R}
\right) U_{\alpha }^{+}\left( \mathbf{P,R}\right) =\varepsilon _{\alpha
}\left( \mathbf{P,R}\right) \text{ if }\left[ \mathbf{P,R}\right] =-i\alpha
\end{equation}
and similarly for $\alpha +d\alpha $. 
\begin{equation}
U_{\alpha +d\alpha }\left( \mathbf{P},\mathbf{R}\right) H_{0}\left( \mathbf{
P,R}\right) U_{\alpha +d\alpha }^{+}\left( \mathbf{P,R}\right) =\varepsilon
_{\alpha +d\alpha }\left( \mathbf{P,R}\right) \text{ if }\left[ \mathbf{P,R}
\right] =-i\left( \alpha +d\alpha \right)
\end{equation}
Let us develop this last relation to the first order in $d\alpha $, 
\begin{equation}
\varepsilon _{\alpha +d\alpha }\left( \mathbf{P,R}\right) =U_{\alpha
}H_{0}U_{\alpha }^{+}+d\alpha \left( \partial _{\alpha }U_{\alpha
}H_{0}U_{\alpha }^{+}+U_{\alpha }H_{0}\partial _{\alpha }U_{\alpha
}^{+}\right)
\end{equation}%
After rewriting the r.h.s. of this equation in terms of Berry connections $
\mathcal{A}_{\alpha }^{R_{l}}=iU_{\alpha }\nabla _{P_{l}}U_{\alpha }^{+}$
and $\mathcal{A}_{\alpha }^{P_{l}}=-iU_{\alpha }\nabla _{R_{l}}U_{\alpha
}^{+}$ we arrive at the following differential equation \cite{SERIESPIERRE} :

\begin{eqnarray}
\frac{d}{d\alpha }\varepsilon _{\alpha }\left( \mathbf{P,R}\right) &=&\left[
\partial _{\alpha }U_{\alpha }\left( \mathbf{P},\mathbf{R}\right) U_{\alpha
}^{+}\left( \mathbf{P,R}\right) ,\varepsilon _{\alpha }\left( \mathbf{P,R}
\right) \right]  \notag \\
&&+\left\{ \frac{1}{2}\mathcal{A}_{\alpha }^{R_{l}}\nabla
_{R_{l}}\varepsilon _{\alpha }\left( \mathbf{P,R}\right) +\nabla
_{R_{l}}\varepsilon _{\alpha }\left( \mathbf{P,R}\right) \mathcal{A}_{\alpha
}^{R_{l}}+\mathcal{A}_{\alpha }^{P_{l}}\nabla _{P_{l}}\varepsilon _{\alpha
}\left( \mathbf{P,R}\right) +\nabla _{P_{l}}\varepsilon _{\alpha }\left( 
\mathbf{P,R}\right) \mathcal{A}_{\alpha }^{P_{l}}\right\}  \notag \\
&&+\frac{i}{2}\left\{ \left[ \varepsilon _{\alpha }\left( \mathbf{P,R}
\right) ,\mathcal{A}_{\alpha }^{R_{l}}\right] \mathcal{A}_{\alpha }^{P_{l}}-
\left[ \varepsilon _{\alpha }\left( \mathbf{P,R}\right) ,\mathcal{A}_{\alpha
}^{P_{l}}\right] \mathcal{A}_{\alpha }^{R_{l}}-\left[ \varepsilon _{\alpha
}\left( \mathbf{P,R}\right) ,\left[ \mathcal{A}_{\alpha }^{R_{l}},\mathcal{A}
_{\alpha }^{P_{l}}\right] \right] \right\}  \notag \\
&&+\frac{i}{2}\left\{ Asym\left\{ \nabla _{P_{l}}\nabla _{R_{l}}\varepsilon
_{\alpha }\left( \mathbf{P,R}\right) \right\} -U_{\alpha }Asym\left\{ \nabla
_{P_{l}}\nabla _{R_{l}}H_{0}\left( \mathbf{P,R}\right) \right\} U_{\alpha
}^{+}\right\}  \notag \\
&&-\frac{i}{2}\left[ X\varepsilon _{\alpha }\left( \mathbf{P,R}\right)
-\varepsilon _{\alpha }\left( \mathbf{P,R}\right) X^{+}\right]  \label{DIFF}
\end{eqnarray}
with the notation
\begin{equation}
X=\left( Asym\left[ \nabla _{R^{l}}\nabla _{P_{l}}U_{\alpha }\left( \mathbf{P
},\mathbf{R}\right) \right] \right) U_{\alpha }^{+}\left( \mathbf{P},\mathbf{
R}\right)
\end{equation}
where the linear operation $Asym$ \cite{PIERREGENERAL} acts on a symmetrical
function in $\mathbf{P}$ and $\mathbf{R}$ in the following way : 
\begin{equation}
Asym\left\{ \frac{1}{2}A\left( \mathbf{R}\right) B\left( \mathbf{P}\right) +
\frac{1}{2}B\left( \mathbf{P}\right) A\left( \mathbf{R}\right) \right\} =
\frac{1}{2}\left[ B\left( \mathbf{P}\right) ,A\left( \mathbf{R}\right) 
\right]
\end{equation}
the functions $A\left( \mathbf{R}\right) $ and $B\left( \mathbf{P}\right) $
being typically monomials in $\mathbf{R}$ and $\mathbf{P}$ arising in the
series expansions of the physical quantities.

As shown in \cite{SERIESPIERRE}, we can separate the energy equation Eq. $
\left( \ref{DIFF}\right) $ in a diagonal and a non diagonal part such that
we are led to the following two equations
\begin{eqnarray}
\frac{d}{d\alpha }\varepsilon _{\alpha }\left( \mathbf{P,R}\right) &=&
\mathcal{P}_{+}[\text{R.H.S}.\text{ of Eq. \ref{DIFF}}]  \label{eq1} \\
0 &=&\mathcal{P}_{-}[\text{R.H.S}.\text{ of Eq. \ref{DIFF}}]  \label{eq2}
\end{eqnarray}
These two equations are supplemented by the differential unitarity condition 
\begin{equation}
0=\partial _{\alpha }U_{\alpha }(\mathbf{P},\mathbf{R)}U_{\alpha }^{+}(
\mathbf{P,R)+}U\mathbf{_{\alpha }(\mathbf{P},\mathbf{R)}}\partial _{\alpha }U
\mathbf{_{\alpha }^{+}(\mathbf{P,R)}}-\frac{i}{2}\left( X-X^{+}\right) +
\frac{i}{2}\left[ \mathcal{A}_{\alpha }^{R_{l}},\mathcal{A}_{\alpha }^{P_{l}}%
\right]  \label{eq3}
\end{equation}
The three equations Eqs.$\left( \ref{eq1}\right) $-$\left( \ref{eq3}\right) $
allow to determine recursively in powers of $\alpha $ the energy of the
quantum system in question. Actually, the integration over $\alpha $ of Eq. $
\left( \ref{eq1}\right) $ gives $\varepsilon _{\alpha }\left( \mathbf{P,R}
\right) $ at order $n$ in $\alpha $ when knowing all quantities at order $
n-1 $. By the same token, Eqs.$\left( \ref{eq2}\right) $ and $\left( \ref
{eq3}\right) $ (whose meaning is that $U_{\alpha }\left( \mathbf{P},\mathbf{R
}\right) $ is unitary at each order in $\alpha $) involve $\partial _{\alpha
}U_{\alpha }\left( \mathbf{P},\mathbf{R}\right) $, and allow to recover $
U_{\alpha }\left( \mathbf{P},\mathbf{R}\right) $ at order $n$ by integration
over $\alpha $. As a consequence, the diagonalization process is perfectly
controlled order by order in the series expansion in $\alpha $.

\section{The semiclassical approximation}

In this section we consider the Hamiltonian diagonalization at the
semiclassical level and the resulting equations of motion. Actually, the
semiclassical approximation has recently found new important applications in
particle and solid state physics. Notably, the equations of motion reveal a
new contribution coming from the Berry curvature. This contribution, called
the anomalous velocity, modifies profoundly the dynamics of the particles.
For instance, the spin Hall effect of electrons and holes in semiconductors 
\cite{MURAKAMI1}, as well as the new discovered optical Hall effect \cite{PHOTONPIERRE}
\cite{ALAIN}\cite{BLIOKH2}\cite{MURAKAMI2} can be interpreted
in this context. Similarly, the recent experimental discovery of the
monopole in momentum can also be elegantly interpreted as the influence of
the Berry curvature on the semiclassical dynamics of Bloch electrons \cite{FANG}\cite{ALAINmonopole}.

\subsection{The semiclassical energy}

The consideration of Eq. $\left( \ref{DIFF}\right) $ alone is sufficient to
deduce the semiclassical diagonal Hamiltonian. Indeed, writing $\varepsilon
_{\alpha }=\varepsilon _{0}+\alpha \varepsilon _{1},$ with $\varepsilon _{0}$
the diagonalized energy at the zero order, Eq. $\left( \ref{DIFF}\right) $
is solved by (putting $\alpha =\hbar $) 
\begin{eqnarray}
\varepsilon \left( \mathbf{P,R}\right) & =\varepsilon _{0}\left( \mathbf{P,R}
\right) +\hbar \left\{ \frac{1}{2}\emph{A}_{0}^{R_{l}}\nabla
_{R_{l}}\varepsilon _{0}\left( \mathbf{P,R}\right) +\nabla
_{R_{l}}\varepsilon _{0}\left( \mathbf{P,R}\right) \emph{A}_{0}^{R_{l}}+
\emph{A}_{0}^{P_{l}}\nabla _{P_{l}}\varepsilon _{0}\left( \mathbf{P,R}
\right) \right.  \notag \\
& +\left. \nabla _{P_{l}}\varepsilon _{0}\left( \mathbf{P,R}\right) \emph{A}
_{0}^{P_{l}}\right\} +\frac{i\hbar }{2}\mathcal{P}_{+}\left\{ \left[
\varepsilon _{0}\left( \mathbf{P,R}\right) ,\mathcal{A}_{0}^{R_{l}}\right] 
\mathcal{A}_{0}^{P_{l}}-\left[ \varepsilon _{0}\left( \mathbf{P,R}\right) ,
\mathcal{A}_{0}^{P_{l}}\right] \mathcal{A}_{0}^{R_{l}}\right\}
\label{semiclassicalenergy}
\end{eqnarray}
where we have introduced the notations $\emph{A}_{0}^{\mathbf{R}}=\mathcal{P}
_{+}\left[ \mathcal{A}_{0}^{\mathbf{R}}\right] $ and $\emph{A}_{0}^{\mathbf{P
}}=\mathcal{P}_{+}\left[ \mathcal{A}_{0}^{\mathbf{P}}\right] $.

This latter expression can also be written 
\begin{equation}
\varepsilon \left( \mathbf{p,r}\right) \simeq \varepsilon _{0}\left( \mathbf{%
p,r}\right) +\frac{i\hbar }{2}\mathcal{P}_{+}\left[ \left[ \varepsilon
_{0}\left( \mathbf{p,r}\right) ,\mathcal{A}_{0}^{R_{l}}\right] \mathcal{A}%
_{0}^{P_{l}}-\left[ \varepsilon _{0}\left( \mathbf{p,r}\right) ,\mathcal{A}%
_{0}^{P_{l}}\right] \mathcal{A}_{0}^{R_{l}}\right] +O(\hbar ^{2})
\label{eqenersc}
\end{equation}%
where we have defined the projected dynamical operators (covariant
coordinates and momentum operators)%
\begin{eqnarray}
\mathbf{r} &=&\mathbf{R}+\hbar \emph{A}_{0}^{\mathbf{R}}  \notag \\
\mathbf{p} &=&\mathbf{P}+\hbar \emph{A}_{0}^{\mathbf{P}}
\end{eqnarray}
with $\mathcal{A}_{0}^{\mathbf{R}}=i\left[ U_{0}\nabla _{\mathbf{P}}U_{0}^{+}%
\right] $, $\mathcal{A}_{0}^{\mathbf{P}}=-i\left[ U_{0}\nabla _{\mathbf{R}%
}U_{0}^{+}\right] ,$and $\mathcal{A}_{0}^{\mathbf{P,R}}=\left[ \nabla _{%
\mathbf{R}}\nabla _{\mathbf{P}}U_{0}\right] U_{0}^{+}$.

The matrix $U_{0}\left( \mathbf{P,R}\right) $ is the diagonalization matrix
for $H_{0}$ when the operators are supposed to be commuting quantities, the
diagonalized energy being $\varepsilon_{0}\left( \mathbf{P,R}\right) .$ When 
$\mathbf{P}$ and $\mathbf{R}$ do not commute, the matrix $U_{0}\left( 
\mathbf{P,R}\right) $ does not diagonalize $H_{0}$ anymore. In order to get
the corrections to the energy at the semiclassical order due to the
noncommutativity of $\mathbf{P}$ and $\mathbf{R}$ we have to compute $%
\varepsilon_{1}\left( \mathbf{P,R}\right) $. From the diagonal Hamiltonian,
we can now derive the equations of motion for the covariant operators.

\subsection{The equations of motion}

Given the Hamiltonian derived in the previous subsection, the equations of
motion can now be easily derived. The evolution equations have to be
considered, not for the usual position and momentum, but rather for the
projected variables $\mathbf{r}$ and $\mathbf{p}$ . Actually, these latter
naturally appear in our diagonalization process at the $\hbar $ order. Let
us remark that their components do not commute any more. Actually%
\begin{eqnarray}
\left[ r_{i},r_{j}\right] &=&i\hbar ^{2}\Theta _{ij}^{rr}=i\hbar ^{2}\left(
\nabla _{P_{i}}\emph{A}_{R_{j}}-\nabla _{P_{j}}\emph{A}_{R_{i}}\right)
+\hbar ^{2}\left[ \emph{A}_{R_{j}},\emph{A}_{R_{i}}\right]  \notag \\
\left[ p_{i},p_{j}\right] &=&i\hbar ^{2}\Theta _{ij}^{pp}=-i\hbar ^{2}\left(
\nabla _{R_{i}}\emph{A}_{P_{j}}-\nabla _{R_{j}}\emph{A}_{P_{i}}\right)
+\hbar ^{2}\left[ \emph{A}_{P_{i}},\emph{A}_{P_{j}}\right]  \notag \\
\left[ p_{i},r_{j}\right] &=&-i\hbar \delta _{ij}+i\hbar ^{2}\Theta
_{ij}^{pr}=-i\hbar \delta _{ij}-i\hbar ^{2}\left( \nabla _{R_{i}}\emph{A}%
_{R_{j}}+\nabla _{P_{j}}\emph{A}_{P_{i}}\right) +\hbar ^{2}\left[ \emph{A}%
_{P_{i}},\emph{A}_{R_{j}}\right]
\end{eqnarray}
the $\Theta _{ij}$ being the so called Berry curvatures.

Using now our Hamiltonian yields directly to general equations of motion for 
$\mathbf{r}$, $\mathbf{p}$ :%
\begin{eqnarray}
\mathbf{\dot{r}} &=&\frac{i}{\hbar }\left[ \mathbf{r},\varepsilon \left( 
\mathbf{p,r}\right) \right] +\frac{i}{\hbar }\left[ \mathbf{r},\frac{i\hbar 
}{2}\mathcal{P}_{+}\left[ \left[ \varepsilon \left( \mathbf{p,r}\right) ,%
\emph{A}_{R_{l}}\right] \emph{A}_{P_{l}}-\left[ \varepsilon \left( \mathbf{%
p,r}\right) ,\emph{A}_{P_{l}}\right] \emph{A}_{R_{l}}\right] \right]  \notag
\\
\mathbf{\dot{p}} &=&\frac{i}{\hbar }\left[ \mathbf{p},\varepsilon \left( 
\mathbf{p,r}\right) \right] +\frac{i}{\hbar }\left[ \mathbf{p},\frac{i\hbar 
}{2}\mathcal{P}_{+}\left[ \left[ \varepsilon \left( \mathbf{p,r}\right) ,%
\emph{A}_{R_{l}}\right] \emph{A}_{P_{l}}-\left[ \varepsilon \left( \mathbf{%
p,r}\right) ,\emph{A}_{P_{l}}\right] \emph{A}_{R_{l}}\right] \right]
\end{eqnarray}
The commutators can be computed through the previous commutation rules
between $\mathbf{r}$ and $\mathbf{p}$. The last term in each equation
represents a contribution of \textquotedblright
magnetization\textquotedblright\ type and has the advantage to present this
general form whatever the system initially considered. In the context of
Bloch electrons in a magnetic field, it gives exactly the magnetization term
revealed in \cite{NIU1} (see \cite{PIERREbloch} and below). For spinning
particles in static gravitational fields, this term gives a coupling between
the spin and the intrinsic angular momentum with magneto-torsion fields \cite%
{PHOTONPIERRE}\cite{PIERREelectrongravit}.

\bigskip

\section{Physical Applications}

\subsection{Electron in a magnetic bloch band}

This topic was first dealt with in \cite{NIU1} in the context of wave
packets dynamics. In the context of the Hamiltonian diagonalization it was
considered in \cite{PIERREbloch} and \cite{PIERREGENERAL}. The purpose is to
find the semiclassical diagonal energy operator for an electron in a
periodic potential facing an electromagnetic field. To apply our formalism,
consider an electron in a crystal lattice perturbated by the presence of an
external electromagnetic field. As is usual, we express the total magnetic
field as the sum of a constant field $\mathbf{B}$ and small nonuniform part $%
\delta \mathbf{B}(\mathbf{R})$. The Schr$\overset{..}{\text{o}}$dinger
equation reads $\left( H_{0}-e\phi(\mathbf{R})\right) \Psi(\mathbf{R})=E\Psi(%
\mathbf{R})$ with $H_{0}$ the magnetic contribution ($\phi$ being the
electric potential) which reads 
\begin{equation}
H_{0}=\left( \frac{\mathbf{P}}{2m}+e\mathbf{A}(\mathbf{R})+e\delta \mathbf{A}%
(\mathbf{R})\right) ^{2}+V(\mathbf{R}),\text{ \ \ \ \ }\mathbf{P=-}%
i\hbar\nabla  \label{Hmagnetic}
\end{equation}
where $\mathbf{A}(\mathbf{R})$ and $\delta\mathbf{A}(\mathbf{R})$ are the
vectors potential of the homogeneous and inhomogeneous magnetic field,
respectively, and $V(\mathbf{R})$ the periodic potential. The large constant
part $\mathbf{B}$ is chosen such that the magnetic flux through a unit cell
is a rational fraction of the flux quantum $h/e$. The advantage of such a
decomposition is that for $\delta\mathbf{A}(\mathbf{R})=0$ the magnetic
translation operators $\mathbf{T}(\mathbf{b})=\exp(i\mathbf{K}.\mathbf{b})$,
with $\mathbf{K}$ the generator of translation, are commuting quantities
allowing to exactly diagonalize the Hamiltonian and to treat $\delta \mathbf{%
A}(\mathbf{R})$ as a small perturbation. The state space of the Bloch
electron in the periodic zone scheme\textbf{\ }\cite{KITTEL} is spanned by
the basis vectors of plane waves\textbf{\ }$\left\vert n,\mathbf{k}%
\right\rangle =\left\vert \mathbf{k}\right\rangle \otimes\left\vert
n\right\rangle $ with $n$ corresponding to a band index and $\mathbf{k}$
vary in $R^{3}$. The state $\left\vert n\right\rangle $ can be seen as a
canonical base vector $\left\vert n\right\rangle =(0...010...0...)$ (with $1$
at the $n$th position) such that $U^{+}\left( \mathbf{k}\right) \left\vert
n\right\rangle =\left\vert u_{n}\left( \mathbf{k}\right) \right\rangle $
with $\left\vert u_{n}\left( \mathbf{k}\right) \right\rangle $ the periodic
part \textbf{(}in space\textbf{)} of the magnetic Bloch waves \cite{NIU1}%
\cite{LANDAU}. In this representation $\mathbf{K}\left\vert n,\mathbf{k}%
\right\rangle =\mathbf{k}\left\vert n,\mathbf{k}\right\rangle $ and
consequently\textbf{\ }the position operator is $\mathbf{R=}%
i\partial/\partial\mathbf{k}$, implying the canonical commutation relations $%
\left[ \mathbf{R}_{i}\mathbf{,K}_{j}\right] =i\delta_{ij}$.

We first perform the diagonalization of the Hamiltonian in Eq. $\left( \ref%
{Hmagnetic}\right) $ for $\delta\mathbf{A}=0$ by diagonalizing
simultaneously $H_{0}$ and the magnetic translation operators $\mathbf{T}$.
The diagonalization is performed as follows: start with an arbitrary basis
of eigenvectors of $\mathbf{T}$. In this basis $H_{0}$ can be seen as a
square matrix with operators entries.$H_{0}$ is diagonalized through a
unitary matrix $U(\mathbf{K})$ which should depend only on $\mathbf{K}$
(since $U$ should leave $\mathbf{K}$ invariant, i.e., $U\mathbf{K}U^{+}=%
\mathbf{K}$) and whose precise expression is not necessary for the
derivation of the equations of motion, such that $UHU^{+}=\mathcal{E}(%
\mathbf{K})-e\phi(U\mathbf{R}U^{+})$, where $\mathcal{E}(\mathbf{K})$ is the
diagonal energy matrix made of elements $\mathcal{E}_{n}(\mathbf{K})$ with $%
n $ the band index (i.e. the diagonal representation of $H_{0}$).

Now, to add a perturbation $\delta A(\mathbf{R)}$ as in (\cite{PIERREbloch}%
), that breaks the translational symmetry, we have to replace $\mathbf{K}$
in all expressions by 
\begin{equation}
\mathbf{\tilde{K}}=\mathbf{K}+e\frac{\delta A(\mathbf{R)}}{\hbar}
\end{equation}
and as the flux $\mathbf{\delta B}$ on a plaquette is not a rational
multiple of the flux quantum, we cannot diagonalize simultaneously its
components $\tilde{K}_{i}$ since they do not commute anymore. Actually 
\begin{equation}
\hbar\lbrack\tilde{K}^{i},\tilde{K}^{j}]=-ie\varepsilon^{ijk}\delta B_{k}(%
\mathbf{R})
\end{equation}
To do the semiclassical diagonalization we replace $U(\mathbf{K})$ by $%
U\left( \mathbf{\tilde{K}}\right) $, so that the non projected Berry
connections are $\mathcal{A}_{R_{i}}=iU\nabla_{\widetilde{K}_{i}}U^{+}$ and $%
\mathcal{A}_{K_{l}}=\nabla_{R_{l}}\delta A_{k}\mathbf{(R)}\mathcal{A}%
_{R_{k}}.$ From these we can define the $n$th intraband position and
momentum operators $\mathbf{r}_{n}\mathbf{=R+}\emph{A}_{n}$ and $\mathbf{%
\tilde{k}}_{n}\simeq\tilde{\mathbf{K}}-e\emph{A}_{n}(\tilde{\mathbf{k}}%
_{n})\times \delta\mathbf{B}(\mathbf{r}_{n})/\hbar+O(\hbar)$ with $\emph{A}%
_{n}=\emph{P}_{n}(U\nabla_{\widetilde{\mathbf{K}}}U^{+})$ the projection of
the Berry connection on the chosen $n$th Band \cite{PIERREbloch}. It can be
readily seen that the matrix elements of\textbf{\ }$\emph{A}_{n}$\textbf{\ }%
can be written $\emph{A}_{n}\left( \mathbf{k}\right) =i\left\langle
u_{n}\left( \mathbf{k}\right) \right\vert \nabla_{\mathbf{k}}\left\vert
u_{n}\left( \mathbf{k}\right) \right\rangle $\textbf{\ (}see also ref. \cite%
{LANDAU}for the derivation of the position operator in the diagonal
representation). What is totally new here is the transformation on the
momentum operator $\tilde{k}_{n}$\ which get also a Berry connection
correction.

Using our general results of section II, the full Hamiltonian Eq. $\left( %
\ref{Hmagnetic}\right) $ can thus be diagonalized through the transformation 
$U(\mathbf{\tilde{K}})+\frac{i}{4\hbar }\left[ \mathcal{A}_{R_{l}},\mathcal{A%
}_{P^{l}}\right] U(\mathbf{\tilde{K}})$ plus a projection on the chosen $n$%
-th Band as it is usual in solid state physics (the so called one band
approximation) and we obtain the energy operator of the $n-$th band as%
\begin{eqnarray}
\emph{P}_{n}\left[ U\left( \mathbf{\tilde{K}}\right) HU^{+}\left( \mathbf{%
\tilde{K}}\right) \right] &=&\emph{P}_{n}\left[ \mathcal{E}\left( \mathbf{%
\tilde{k}}\right) -\frac{i}{4}\left[ \mathcal{E}(\mathbf{K}),U\nabla
_{K_{i}}U^{+}\right] \varepsilon ^{ijk}\frac{\delta B^{k}(\mathbf{r)}}{\hbar 
}U\nabla _{K_{j}}U^{+}\right.  \notag \\
&&\left. -\frac{i}{4}U\nabla _{K_{j}}U^{+}\left[ \mathcal{E}(\mathbf{K}%
),U\nabla _{K_{i}}U^{+}\right] \varepsilon ^{ijk}\frac{\delta B^{k}(\mathbf{%
r)}}{\hbar }\right]  \notag \\
&=&\mathcal{E}_{n}\left( \mathbf{\tilde{k}}_{n}\right) -\mathcal{M}(\tilde{%
\mathbf{K}}).\delta \mathbf{B}(\mathbf{\mathbf{r}}_{n}\mathbf{)+}O\mathbf{(}%
\hbar ^{2}\mathbf{)}  \label{ENER}
\end{eqnarray}
where the energy levels $\mathcal{E}_{n}\left( \mathbf{\tilde{k}}_{n}\right) 
$ are the same as $\mathcal{E}_{n}(\mathbf{K})$ with $\mathbf{\tilde{k}}_{n}$
replacing $\mathbf{K.}$ The magnetization $\mathcal{M}(\tilde{\mathbf{K}})=%
\emph{P}_{n}(\frac{ie}{2\hbar }\left[ \mathcal{E}(\tilde{\mathbf{K}}),%
\mathcal{A}(\tilde{\mathbf{K}})\right] \times \mathcal{A}(\tilde{\mathbf{K}}%
))$ can be written under the usual form \cite{LANDAU} in the $(\mathbf{k},n)$
representation 
\begin{equation}
\mathcal{M}_{nn}^{i}=\frac{ie}{2\hbar }\varepsilon ^{ijk}\sum_{n^{\prime
}\neq n}(\mathcal{E}_{n}-\mathcal{E}_{n^{\prime }})(\mathcal{A}%
_{j})_{nn^{\prime }}(\mathcal{A}_{k})_{n^{\prime }n}
\end{equation}%
We mention that this magnetization (the orbital magnetic moment of Bloch
electrons), has been obtained previously in the context of electron wave
packets dynamics \cite{NIU1}.

From the expression of the energy Eq. $\left( \ref{ENER}\right) $ we can
deduce the equations of motion (with the band index $n$ now omitted) 
\begin{eqnarray}
\dot{\mathbf{r}} &=&\partial E(\tilde{\mathbf{k}})/\hbar \partial \tilde{
\mathbf{k}}-\dot{\tilde{\mathbf{k}}}\times \Theta (\tilde{\mathbf{k}}) 
\notag \\
\hbar \dot{\tilde{\mathbf{k}}} &=&-e\mathbf{E}-e\dot{\mathbf{r}}\times
\delta \mathbf{B}(\mathbf{r})-\mathcal{\mathbf{M}}\partial \delta \mathbf{B}/\partial \mathbf{r}  \label{EQM}
\end{eqnarray}
where $\left[ r^{i},r^{j}\right] =i\Theta ^{ij}(\widetilde{\mathbf{k}})$
with $\Theta ^{ij}(\widetilde{\mathbf{k}})=\partial ^{i}\mathcal{A}^{j}(%
\widetilde{\mathbf{k}})-\partial ^{j}\mathcal{A}^{i}(\widetilde{\mathbf{k}})$
the Berry curvature. As explained in \cite{PIERREbloch} these equations are
the same as the one derived in \cite{NIU1} from a completely different
formalism.

\subsection{Photon in a static gravitational field.}

We now apply our general approach to the case of a photon propagating in an
arbitrary static gravitational field, where $g_{0i}=0$ for $i=1,2,3$, so
that $ds^{2}=g_{00}(dx^{0})^{2}-g_{ij}dx^{i}dx^{j}=0$. As explained in \cite%
{PHOTONPIERRE} the photon description is obtained by considering first a
Dirac massless particle (massless neutrino) and then by replacing the Pauli
matrices $\mathbf{\sigma }$ by the spin-$1$ matrices $\mathbf{S.}$ Therefore
we start with the Dirac Hamiltonian in static gravitational field which can
be written 
\begin{equation}
\hat{H}=\sqrt{g_{00}}\mathbf{\alpha }.\mathbf{\tilde{P}}+\frac{\hbar }{4}%
\varepsilon _{\varrho \beta \gamma }\Gamma _{0}^{\varrho \beta }\sigma
^{\gamma }+i\frac{\hbar }{4}\Gamma _{0}^{0\beta }\alpha _{\beta }
\label{HGR}
\end{equation}
with $\mathbf{\tilde{P}}$ given by $\tilde{P}_{\alpha }\mathbf{=}h_{\alpha
}^{i}(\mathbf{R})(P_{i}+\frac{\hbar }{4}\varepsilon _{\varrho \beta \gamma
}\Gamma _{i}^{\varrho \beta }\sigma ^{\gamma })$ with $h_{\alpha }^{i}$ the
static orthonormal dreibein $(\alpha =1,2,3)$, $\Gamma _{i}^{\alpha \beta }$
the spin connection components and $\varepsilon _{\alpha \beta \gamma
}\sigma ^{\gamma }=\frac{i}{8}(\gamma ^{\alpha }\gamma ^{\beta }-\gamma
^{\beta }\gamma ^{\alpha }).$ The coordinate operator is again given by $%
\mathbf{R=}i\hbar \partial _{\mathbf{p}}.$ Note that here we consider the
general case where an arbitrary static torsion of space is allowed. It is
known \cite{LECLERC} that for a static gravitational field (which is the
case considered here), the Hamiltonian $\hat{H}$ is Hermitian. We now want
to diagonalize $\hat{H}$ through a unitary transformation $U(\mathbf{\tilde{P%
}}).$ Because the components of $\mathbf{\tilde{P}}$ depend both on
operators $\mathbf{P}$ and $\mathbf{R}$ the diagonalization at order $\hbar $
is performed by adapting the method detailed above to block-diagonal
Hamiltonians. To do so, we first write $\hat{H}$ in a symmetrical way in $%
\mathbf{P}$ and $\mathbf{R}$ at first order in $\hbar $. This is easily
achieved using the Hermiticity of the Hamiltonian which yields \ 
\begin{equation}
\hat{H}=\frac{1}{2}\left( \sqrt{g_{00}}\mathbf{\alpha }.\mathbf{\tilde{P}+%
\tilde{P}}^{+}\mathbf{.\alpha }\sqrt{g_{00}}\right) +\frac{\hbar }{4}%
\varepsilon _{\varrho \beta \gamma }\Gamma _{0}^{\varrho \beta }\sigma
^{\gamma }.
\end{equation}
Using the general expression Eq. $\left( \ref{eqenersc}\right) $ we arrive
at the following expression for the diagonal positive (we have projected on
the positive energy subspace) energy representation $\tilde{\varepsilon}:$ 
\begin{equation}
\tilde{\varepsilon}=\varepsilon +\frac{\lambda }{4}\frac{\mathbf{p.\Gamma }%
_{0}}{p}+\frac{\hbar \mathbf{B}.\mathbf{\sigma }}{2\varepsilon }-\frac{(%
\mathbf{A}_{R}\mathbf{\times p).B}}{\varepsilon ^{(r)}}.  \label{energy}
\end{equation}
where we have introduced a field $B_{\gamma }=-\frac{1}{2}P_{\delta
}T^{\alpha \beta \delta }\varepsilon _{\alpha \beta \gamma }$ with $%
T^{\alpha \beta \delta }=h_{k}^{\delta }\left( h^{l\alpha }\partial
_{l}h^{k\beta }-h^{l\beta }\partial _{l}h^{k\alpha }\right) +h^{l\alpha
}\Gamma _{l}^{\beta \delta }-h^{l\beta }\Gamma _{l}^{\alpha \delta }$ the
usual torsion for a static metric (where only space indices in the
summations give non zero contributions). We have also defined in Eq. $\left( %
\ref{energy}\right) $ 
\begin{equation}
\varepsilon =c\sqrt{\left( p_{i}+\frac{\lambda }{4}\frac{\Gamma _{i}(\mathbf{%
r}).\mathbf{p}}{p}\right) g^{ij}g_{00}\left( p_{j}+\frac{\lambda }{4}\frac{%
\Gamma _{j}(\mathbf{r}).\mathbf{p}}{p}\right) },
\end{equation}%
with the $\gamma $-th component of the vector $\mathbf{\Gamma }_{i}$ as $%
\Gamma _{i,\gamma }=\varepsilon _{\varrho \beta \gamma }\Gamma _{i}^{\varrho
\beta }(\mathbf{r})$ and the helicity $\lambda =\frac{\hbar \mathbf{p.\sigma 
}}{p}.$ Note that the dynamical operators are now
\begin{eqnarray}
\mathbf{r} &=&\mathbf{R}+\hbar c^{2}\frac{\mathbf{P}\times \mathbf{\Sigma }}{%
2\varepsilon ^{2}}  \label{r} \\
\mathbf{p} &=&\mathbf{P}-\hbar c^{2}(\frac{\mathbf{P}\times \mathbf{\Sigma }%
}{2\varepsilon ^{2}})\mathbf{\nabla }_{\mathbf{R}}\mathbf{\tilde{P}}
\label{p}
\end{eqnarray}

Interestingly, this semi-classical Hamiltonian presents formally the same
form as the one of a Dirac particle in a true external magnetic field \cite%
{PIERREGENERAL}\cite{BLIOKH1}. The term $\mathbf{B}.\mathbf{\sigma }$ is
responsible for the Stern-Gerlach effect, and the operator $\mathbf{L}=(%
\mathbf{A}_{R}\mathbf{\times p)}$ is the intrinsic angular momentum of
semiclassical particles. The same contribution appears also in the context
of the semiclassical behavior of Bloch electrons (spinless) in an external
magnetic field \cite{PIERREbloch}\cite{NIU1} where it corresponds to a
magnetization term. Because of this analogy and since $T^{\alpha \beta
\delta }$ is directly related to the torsion of space through $T^{\alpha
\beta \delta }=h_{k}^{\delta }h^{i\alpha }h^{j\beta }T_{ij}^{k}$ we call $%
\mathbf{B}$ a magnetotorsion field.

However, this form for the energy presents the default to involve the spin
rather than the helicity. Actually one can use the property $\lambda \mathbf{%
p}/2p=\hbar\mathbf{\sigma/}2-(\mathbf{A}_{R}\mathbf{\times p)}$ to rewrite
the energy as 
\begin{equation}
\widetilde{\varepsilon}\simeq\varepsilon+\frac{\lambda}{4}\frac {\mathbf{%
p.\Gamma}_{0}}{p}+\frac{\lambda g_{00}}{2\varepsilon}\frac {\mathbf{B}.%
\mathbf{p}}{p}  \label{NUMBER}
\end{equation}

The semi-classical Hamiltonian Eq. $\left( \ref{NUMBER}\right) $ contains,
in addition to the energy term $\varepsilon$, new contributions due to the
Berry connections. Indeed, Eq. $\left( \ref{NUMBER}\right) $ shows that the
helicity couples to the gravitational field through the magnetotorsion field 
$\mathbf{B}$ which is non-zero for a space with torsion. As a consequence, a
hypothetical torsion of space may be revealed through the presence of this
coupling. Note that, in agrement with \cite{SILENKOgravit}, this Hamiltonian
does not contain the spin-gravity coupling term $\mathbf{\Sigma.\nabla}%
g_{00} $ predicted in \cite{OBUKHOV}.

From Eqs. $\left( \ref{r}\right) $ and $\left( \ref{p}\right) $ we deduce
the new (non-canonical) commutations rules%
\begin{eqnarray}
\left[ r^{i},r^{j}\right] &=&i\hbar \Theta _{rr}^{ij}  \notag \\
\left[ p^{i},p^{j}\right] &=&i\hbar \Theta _{pp}^{ij}  \notag \\
\left[ p^{i},r^{j}\right] &=&-i\hbar g^{ij}+i\hbar \Theta _{pr}^{ij}
\end{eqnarray}
where $\Theta _{\zeta \eta }^{ij}=\partial _{\zeta ^{i}}A_{\eta
^{j}}-\partial _{\eta ^{i}}A_{\zeta ^{j}}+[A_{\zeta ^{i},}A_{\eta ^{j}}]$
where $\zeta ,\eta $.mean either $r$ or $p$. An explicit computation shows
that at leading order%
\begin{eqnarray}
\Theta _{rr}^{ij} &=&-\hbar c^{4}\frac{\left( \mathbf{\Sigma }.\mathbf{p}%
\right) p_{\gamma }}{2\varepsilon ^{4}}\varepsilon ^{\alpha \beta \gamma
}h_{\alpha }^{i}h_{\beta }^{j}  \notag \\
\Theta _{pp}^{ij} &=&-\hbar c^{4}\frac{\left( \mathbf{\Sigma }.\mathbf{p}%
\right) p_{\gamma }}{2\varepsilon ^{4}}\nabla _{r_{i}}p_{\alpha }\nabla
_{r_{j}}p_{\beta }\varepsilon ^{\alpha \beta \gamma }  \notag \\
\Theta _{pr}^{ij} &=&\hbar c^{4}\frac{\left( \mathbf{\Sigma }.\mathbf{p}%
\right) p_{\gamma }}{2\varepsilon ^{4}}\nabla _{r_{i}}p_{\alpha }h_{\beta
}^{j}\varepsilon ^{\alpha \beta \gamma }
\end{eqnarray}
From the additional commutation relations between the helicity and the
dynamical operators $[r_{i},\lambda ]=[p_{i},\lambda ]=0$ we deduce the
semiclassical equations of motion%
\begin{eqnarray}
\mathbf{\dot{r}} &=&\left( 1-\Theta _{pr}\right) \nabla _{\mathbf{p}}\tilde{%
\varepsilon}+\mathbf{\dot{p}\times }\Theta _{rr}  \notag \\
\mathbf{\dot{p}} &=&-\left( 1-\Theta _{pr}\right) \nabla _{\mathbf{r}}\tilde{%
\varepsilon}+\mathbf{\dot{r}}\times \Theta _{pp}  \label{Eqmotion}
\end{eqnarray}
To complete the dynamical description of the photon notice that at the
leading order the helicity $\lambda $ is not changed by the unitary
transformation which diagonalizes the Hamiltonian so that it can be written $%
\lambda =\hbar \mathbf{p}.\mathbf{\Sigma }/p$. After a short computation one
can check that the helicity is always conserved 
\begin{equation}
\frac{d}{dt}\left( \frac{\hbar \mathbf{p.\Sigma }}{p}\right) =0
\end{equation}
for an arbitrary static gravitational field independently of the existence
of a torsion of space.

Eqs. $\left( \ref{Eqmotion}\right) $ are the new semiclassical equations of
motion for a photon in a static gravitational field. They describe the ray
trajectory of light in the first approximation of geometrical optics (GO).
(In GO it is common to work with dimensionless momentum operator $\mathbf{p=}%
k_{0}^{-1}\mathbf{k}$ with $k_{0}=\omega/c$ instead of the momentum \cite%
{BLIOKH2}). For zero Berry curvatures we obtain the well known zero order
approximation of GO and photons follow the null geodesic. The velocity
equation contains the by now well known anomalous contribution $\mathbf{%
\dot {p}\times}\Theta^{rr}$ which is at the origin of the intrinsic spin
Hall effect (or Magnus effect) of the photon in an isotropic inhomogeneous
medium of refractive index $n(r)$ \cite{ALAIN}\cite{BLIOKH2}\cite{MURAKAMI2}\cite{HORVATHY3}.
Indeed, this term causes an additional displacement of photons of distinct
helicity in opposite directions orthogonally to the ray. Consequently, we
predict gravitational birefringence since photons with distinct helicities
follow different geodesics. In comparison to the usual velocity $\mathbf{%
\dot{r}}=\nabla_{\mathbf{p}}\tilde{\varepsilon}$ $\sim c$, the anomalous
velocity term $\mathbf{v}_{\perp}$ is obviously small, its order $%
v_{\perp}^{i}\sim c\widetilde{\lambda}\nabla_{r^{j}}g^{ij}$ being
proportional to the wave length $\widetilde{\lambda}$.

The momentum equation presents the dual expression $\mathbf{\dot{r}}%
\times\Theta_{pp}$ of the anomalous velocity which is a kind of Lorentz
force which being of order $\hbar$ does not influence the velocity equation
at order $\hbar$. Note that similar equations of motion with dual
contributions $\mathbf{\dot{p}\times}\Theta_{rr}$ and $\mathbf{\dot{r}}
\times\Theta_{pp}$ were predicted for the semiclassical dynamics of spinless
electrons in crystals subject to small perturbations \cite{NIU1}\cite{PIERREbloch}.

\bigskip

\section{Conclusion}

Some recent applications of semiclassical methods to several branches of
Physics, such as spintronics or solid state physics have shown the relevance
of Berry Phases contributions to the dynamics of a system. However, these
progresses called for a rigorous Hamiltonian treatment that would allow for
deriving naturally the role of the Berry phase. In this paper we have
considered a diagonalization method for a broad class of quantum systems,
including the electron in a periodic potential and the Dirac Hamiltonian in
a gravitational field. Doing so, we have exhibited a general pattern for
this class of systems implying the role of the Berry phases both for the
position and the momentum. In such a context, the coordinates and momenta
algebra are no longer commutative, and the dynamical equations for these
variables directly include the influence of Berry phases through the
parameters of noncommutativity (Berry curvatures) and through an abstract
magnetization term.

\textbf{Acknowledgment.} It is a great pleasure for us to thank Prof. S.
Ghosh for collaborations and for having given to H. M. the opportunity to
give a talk at the Kolkata conference on recent developments in theoretical
physics.


\begin{thebibliography}{99}
\bibitem{BERRY} M. V. Berry, Proc. R. Soc. A \textbf{392} (1984) 45.

\bibitem{MURAKAMI1} S. Murakami, N. Nagaosa, S. C. Zhang, Science \textbf{\
301} (2003) 1348

\bibitem{PHOTONPIERRE} P. Gosselin, A. B\'{e}rard, H. Mohrbach, Phys. Rev. D 
\textbf{75} (2007) 084035.

\bibitem{ALAIN} A. B\'{e}rard, H. Mohrbach, Phys. lett. A \textbf{352}
(2006) 190.

\bibitem{NIU1} M. C. Chang, Q. Niu, Phys. Rev. Lett \textbf{75} (1995) 1348;
Phys. Rev. B \textbf{53} (1996) 7010; G. Sundaram, Q. Niu, Phys. Rev. B 
\textbf{59} (1999) 14915.

\bibitem{PIERREbloch} P. Gosselin, F. M\'{e}nas, A. B\'{e}rard, H. Mohrbach,
Europhys. Lett. \textbf{76} (2006) 651.

\bibitem{SUBIRDIRAC} P. Gosselin, A. B\'{e}rard, H. Mohrbach, S. Ghosh,
Phys. Lett. B \textbf{660}\ (2008) 267.

\bibitem{PIERREGENERAL} P. Gosselin, A. B\'{e}rard and H. Mohrbach, Eur.
Phys. J. B \textbf{58}, 137 (2007).

\bibitem{FOLDY} L. L. Foldy and S. A. Wouthuysen, Phys. Rev. \textbf{78}
(1950) 29.

\bibitem{ERIKSEN} E. Eriksen, Phys. Rev. \textbf{111}, 1011 (1958).

\bibitem{NIKITIN} A. G. Nikitin, J. Phys. A \textbf{31}, 3297 (1998).

\bibitem{SILENKO1} A. J. Silenko, J. Math. Phys.\textbf{44}, 2952 (2003).

\bibitem{SERIESPIERRE} P. Gosselin, J. Hanssen and H. Mohrbach, Phys. Rev.
D. in press, arXiv:cond-mat/0611628.

\bibitem{SILENKO} A. J. Silenko, arXiv: math-ph: 0710.4218.

\bibitem{PIERREelectrongravit} P. Gosselin, A. B\'{e}rard, H. Mohrbach,
Phys. Lett. A \textbf{368 }(2007) 356

\bibitem{BLIOKH2} K. Y. Bliokh, Y. P. Bliokh, Phys. Lett A \textbf{333}
(2004), 181; Phys. Rev. E \textbf{70} (2004) 026605; Phys. Rev. Lett. 
\textbf{96} (2006) 073903.

\bibitem{MURAKAMI2} M. Onoda, S. Murakami, N. Nagasoa, Phys. Rev. Lett. 
\textbf{93} (2004) 083901.

\bibitem{FANG} Z. Fang et al., Science \textbf{302}, 92 (2003).

\bibitem{ALAINmonopole} A. B\'{e}rard, H. Mohrbach, Phys. Rev. D \textbf{69}
(2004) 127701

\bibitem{KITTEL} C. Kittel, Quantum Therory of Solids, Wiley, New York,1963..

\bibitem{LANDAU} E. M. Lifshitz, L. P. Pitaevskii, Statistical Physics, vol
9, Pergamon Press, 1981.

\bibitem{LECLERC} M. Leclerc, Class.Quant. Grav. \textbf{23} (2006) 4013.

\bibitem{BLIOKH1} K. Y. Bliokh, Eur. Lett. \textbf{72} (2005) 7.

\bibitem{OBUKHOV} Y. N. Obukhov, Phys. Rev. Lett \textbf{86} (2001) 192.

\bibitem{SILENKOgravit} A. J. Silenko, O. V. Teryaev, Phys. Rev. D \textbf{71%
} (2005) 064016.

\bibitem{HORVATHY3} C. Duval, Z. Horvath, P. Horvathy, L. Martina, P.
Stichel, Phys. Rev. Lett \textbf{96} (2006) 099701; D. Xiao, J. Shi, Q. Niu,
Phys. Rev. Lett. \textbf{96} (2006) 099702.
\end{thebibliography}
\end{document}